\documentclass[10pt]{article}
\usepackage{amsmath,amssymb,amsfonts,amsthm,graphicx,psfig}
\def\pmb#1{\setbox0=\hbox{$#1$}%
  \kern-.025em\copy0\kern-\wd0
  \kern.05em\copy0\kern-\wd0
  \kern-.025em\raise.0433em\box0}
\def\pmbs#1{\setbox0=\hbox{$\scriptstyle #1$}%
  \kern-.0175em\copy0\kern-\wd0
  \kern.035em\copy0\kern-\wd0
  \kern-.0175em\raise.0303em\box0}
\def\be{\begin{equation}}
\def\ee{\end{equation}}
\def\bea{\begin{eqnarray}}
\def\eea{\end{eqnarray}}
\def\lb{\label}
\def\ct{\cite}

\def\rr{k}
\def\bi{\bibitem}
\def\vec#1{\mbox{\boldmath$#1$}}

\def\gam{\gamma}
\def\d{\delta}
\def\eps{\epsilon}

\def\sig{\sigma}

\def\Sig{\Sigma}

\def\Om{\Omega}

\def\udot{\dot{u}}
\def\Udot{\dot{U}}

\def\cn{{\cal N}}
\def\ce{{\cal E}}
\def\ch{{\cal H}}

\def\CC{C_{abcd}C^{abcd}}
\def\CCS{C^*_{abcd}C^{abcd}}
\def\cc{{\cal C}}
\def\ccs{{\cal C}^*}
\def\ccr{C_R}
\def\ccrs{C^*_R}

\def\vece{\vec{e}}

\def\ptl{\partial}
\def\parb{\pmb{\partial}}

\def\hsp5{\hspace{5mm}}
\newcommand{\sfrac}[2]{{\textstyle{#1\over#2}}}
\def\case#1/#2{\textstyle\frac{#1}{#2}}

\def\cqg{{Class.~Quantum~Grav.} }

\begin{document}
\title{{\sc Asymptotic Silence-Breaking Singularities}}
\author{\sc
Woei Chet Lim$^{1}$\thanks{Electronic address: {\tt
wclim@mathstat.dal.ca}} ,
Claes Uggla$^{2}$\thanks{Electronic address:
{\tt claes.uggla@kau.se}}\ and
John Wainwright$^{3}$\thanks{Electronic address: {\tt
jwainwri@math.uwaterloo.ca}} \\
$^{1}${\small\em Department of Mathematics and Statistics, Dalhousie
University, Halifax, Nova Scotia, Canada B3H 3J5}\\
$^{2}${\small\em Department of Physics, University of Karlstad,
S-651 88 Karlstad, Sweden}\\
$^{3}${\small\em Department of Applied Mathematics, University of
Waterloo, Waterloo, Ontario, Canada N2L 3G1}}

\date{\normalsize{February 12, 2006}}
\maketitle
\begin{abstract}

We discuss three complementary aspects of 
scalar curvature singularities: asymptotic causal properties,
asymptotic Ricci and Weyl curvature, and asymptotic spatial
properties. We divide scalar curvature singularities into two
classes: so-called asymptotically silent singularities and
singularities that break asymptotic silence. The
emphasis in this paper is on the latter class which have not been
previously discussed. We illustrate the above aspects and concepts
by describing the singularities of a number of representative
explicit perfect fluid solutions.

\end{abstract}
\centerline{\bigskip\noindent PACS number(s): 04.20.-q, 98.80.Jk,
04.20.Dw, 04.20.Ha~\hfill {gr-qc/0511139}}

\section{Introduction}
\lb{sec:intr}

In a recent paper a general framework for locally analyzing
Einstein's field equations (EFE) was presented~\ct{uggetal03}.
As a first application the
structure of generic spacelike singularities was investigated, and an
attractor describing the asymptotic dynamical properties of such
singularities was presented. The results in \ct{uggetal03} were later
numerically supported in~\cite{gar04} and~\cite{andetal05}.
In~\cite{limetal04} isotropic singularities were investigated by
extending the methods developed in \ct{uggetal03} to also include the null
geodesic equations. Gradually a picture about asymptotic causal
structure has emerged, indicating that generic spacelike singularities and
isotropic singularities are examples in which local null cones
collapse onto a timeline towards the singularity, leading to the
formation of particle horizons with a size that shrinks to zero
(see p. 75 and Fig. 1 on p. 76 in~\cite{hveetal02}). Since the
shrinking of horizons prevents communication, the phenomenon was
referred to as {\em asymptotic silence\/} and a corresponding
singularity as being asymptotically silent (see~\cite{andetal05}
for further discussion). However, for some solutions particle
horizons may not form, or if they do form, they may not shrink to zero 
size. The
best known example showing that the particle horizon can be
broken in one spatial direction is a spatially homogeneous (SH),
Bianchi type I dust solution, with a so-called weak null
singularity (see~\ct{hawell73} p. 176). 
The asymptotic behaviour of this solution towards the
singularity gives one illustration of how 
{\em asymptotic silence-breaking\/} can occur.

Our goal in this paper is to illustrate the notion of asymptotic
silence-breaking 
at a scalar curvature singularity by analyzing a selection of 
explicit perfect fluid solutions,
thereby extending our understanding of the asymptotic causal structure of
singularities.  
Although our examples are non-generic, we believe that they may be of
importance for more general phenomena, worthy of further study.
In this context it is worth pointing out that non-generic
solutions can play an important role in determining the behaviour of 
generic solutions.  For example, self-similar cosmological solutions, 
which are certainly non-generic, have been shown to play an important 
role in determining the dynamics of generic solutions.\footnote{The Kasner
solutions, which are self-similar, determine the past attractor for 
generic cosmological solutions (see \cite{uggetal03}), while the flat FL
solution, 
which is also self-similar, underlies the generic phenomenon of
intermediate isotropization (see \cite{waiell97}, p.\ 312).}

The outline of the paper is as follows. In Sec. 2 we describe three 
complementary geometrical aspects of a matter singularity: firstly,
according to its asymptotic causal properties in terms of
asymptotic silence or asymptotic silence-breaking; secondly,
according to its scalar Ricci and Weyl curvature; thirdly,
according to how a spatial volume element is deformed towards the
singularity. In Sec. 3 we give a number of selected examples that show 
how asymptotic silence can be broken and how this is connected with
the above aspects. In Sec. 4 we conclude with some remarks about
our results and their implications. 
A brief introduction to the orthonormal frame formalism and to 
Hubble-normalized variables is given in Appendix A.
In Appendix B we give the non-tilted SH self-similar solutions,
which describe the asymptotic behaviour of the examples in
Sec. 3. 

Although we make use of the orthonormal frame formalism (see, for example
\cite{ellels99}) for doing the necessary calculations, a detailed
knowledge of this formalism is not required in the main body of the paper.

\section{Geometrical properties of singularities}
\lb{sec:sil}

In this paper we consider perfect fluid
solutions of EFE with a barotropic equation of state $p =
p(\rho)$, where $p$ and $\rho$ are the isotropic pressure and
energy density in the rest frame of the fluid.
Linear equations of state are characterized by the
constant $\gamma$ defined by $p=(\gamma-1)\rho$.
We assume that $\rho>0$ and $0 < \gamma \leq 2$.

The solutions that we consider have a matter singularity, i.e. the
matter density $\rho$ diverges since the overall length scale
$\ell$ tends to zero
\be
\lim_{\ell\rightarrow 0}\,\rho = +\infty\, .
\ee

Our analysis of singularities relies heavily on the use of scale-invariant
variables that are defined by normalizing with the appropriate power
of the Hubble scalar $H$ of the fluid congruence. The Hubble scalar
is defined in terms of the 4-velocity $\mathbf{u}$ of the perfect fluid by
\be
H = \sfrac{1}{3}\,\nabla_a\,u^a\, ,
\ee
and serves to define the length scale $\ell$ according to
\be
H = \frac{\dot\ell}{\ell}\, ,
\ee
where the overdot denotes differentiation along $\bf u$. Important
dimensionless quantities related to $H$ are the 
\emph{deceleration parameter} $q$ and the 
\emph{spatial Hubble gradient} $r_a$, 
defined by
\be 
\label{qr_a}
	q   = -\,\frac{\dot H}{H^2}-1\, ,\quad 
	r_a = -\,\frac{h_a{}^b\,\nabla_b\,H}{H^2}\, .
\ee
where $h_a{}^b=\delta_a{}^b + u_a\,u^b$ projects into the 3-space
orthogonal to $\mathbf{u}$.

We now introduce three complementary geometrical aspects of 
curvature singularities.

\subsection{Asymptotic causal singularity structure}

The first aspect of singularities that we consider is the
asymptotic causal properties towards the singularity. We define a
part of a singularity to be asymptotically silent if all observers
that approach it have particle horizons that shrink to zero size.
If a particle horizon does not shrink to zero, or if a particle
horizon does not form in one or more directions, we say that {\em
asymptotic silence-breaking\/} occurs.

Reference
\ct{uggetal03} heuristically links the formation of shrinking particle 
horizons to the
limit
\be 
\label{E0}
	\lim_{\ell\rightarrow 0}\,E_\alpha{}^i = 0\, 
\ee
(\ct{uggetal03}, p. 103502-10) of the Hubble-normalized components of the
spatial frame vector fields
\be 
\lb{Eai_def}
E_\alpha{}^i = \frac{e_\alpha{}^i}{H}\,  \ee
(these vector fields are introduced in
 Appendix A). However, as we will see, it seems to be
necessary to also demand that
\be 
\lb{Ur0}
	\lim_{\ell\rightarrow 0}\,\Udot_\alpha =0\, ,\quad
	\lim_{\ell\rightarrow 0}\,r_\alpha =0\, ,
\ee
in order to provide sufficient conditions for asymptotic silence to hold.
Here the $r_\alpha$ are the frame components of the spatial Hubble
gradient (\ref{qr_a}), and the $\dot{U}_\alpha$ are the frame components of
the Hubble-normalized acceleration of the fluid (see (\ref{Hnorm}) and
(\ref{qr_alpha}) in the Appendix).

Below we will show how asymptotic silence can be broken in various
ways. We will give examples where
\be\lb{silbreak1}
\lim_{\ell \rightarrow 0} E_\alpha{}^i ={\rm diag}\,(const,0,0)\, ,
\ee
where $const \neq 0$ (if $const$ appears in an asymptotic
expression, it will from now on be assumed to be non-zero),
leading to asymptotic silence-breaking in one direction
only. We refer to such a singularity as {\em partially silent\/}.
It is also possible that
\be \lim_{\ell \rightarrow 0} E_\alpha{}^i =C_\alpha{}^i\, ,
\ee
where the constant $3\times3$ matrix $C_\alpha{}^i$ has
higher rank than 1 (rank 1 yields equation~(\ref{silbreak1})), which
leads to the breaking of asymptotic silence in more than one
direction, although we shall not consider
examples with this property, apart from the
Friedmann-Lema\^{\i}tre
 model to follow.%
\footnote{Asymptotic silence-breaking in more than one direction appears 
to necessitate violation of an energy condition, i.e. $p<0$ or even $\rho 
+3p<0$ (see (\ref{E_FL}) to follow).}
In addition we give examples of asymptotic silence-breaking when
$\lim_{\ell\rightarrow 0}\,E_\alpha{}^i = \infty$
for some values of $\alpha$ and $i$, and when some of the
components of $E_\alpha{}^i$ oscillate indefinitely as $\ell
\rightarrow 0$, so that $\lim_{\ell \rightarrow 0} E_\alpha{}^i$
does not exist.

To illustrate some of the above let us consider the flat
Friedmann-Lema\^{\i}tre
 model:
\bea
ds^2 &=& -dt^2 + t^{4/3\gamma}\,[dr^2 + r^2(d\theta^2 + \sin^2\theta\,d\phi^2)]\, ,\\
\rho &=& \frac{4}{3\gamma^2\,t^2}\, ,\quad p=(\gamma-1)\rho\, ,\quad
{\bf u} = \partial/\partial t\, .
\eea
The radial null geodesics are given by
\be
\left(\frac{dt}{dr}\right)^2 = t^{4/3\gamma}\, .
\ee
Solving for $r=r(t)$ shows that the past light cone at $r=0,t=t_0$ is generated
by the following family of null geodesics;
\be
r-b = \left\{\begin{array}{lll}
    -\,\frac{3\gamma}{3\gamma-2}\, t^{(3\gamma-2)/3\gamma} &,\quad
 b= \frac{3\gamma}{3\gamma-2}\, t_0^{(3\gamma-2)/3\gamma}\, &,\quad
\gamma \neq \sfrac{2}{3}
    \\
    -\,\ln\,t &,\quad  b = \ln\,t_0 &,\quad \gamma = \sfrac{2}{3}
\end{array}\right. \, .
\ee
Observe that
\be
\lim_{t\rightarrow 0^+}\,r =
\left\{\begin{array}{ll}
b &,\quad \sfrac{2}{3} < \gamma \leq 2 \\
+\infty &,\quad 0 < \gamma \leq \sfrac{2}{3}
\end{array}\right. \, .
\ee

\begin{figure}[h]
\begin{center}
\includegraphics{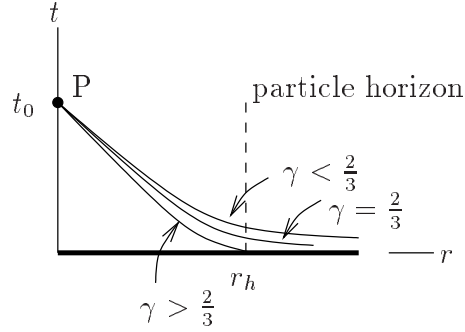}
\caption{The past light cone of the point P and the initial
singularity at $t=0$, showing the particle horizon at $r=r_h$ in
the case $\gamma>\frac{2}{3}$, and the absence of a particle horizon in
the case $\gamma \leq \frac23$.} \label{Fig:part}
\end{center}
\end{figure}

It follows that if $\frac{2}{3} < \gamma \leq  2$, the past light
cone intersects the hypersurface $t=0$ at
\be
	r = r_h(t_0) = \frac{3\gamma}{3\gamma-2} 
			t_0^{(3\gamma-2)/3\gamma}\, ,
\ee
thereby creating a particle horizon, as shown in Fig.~\ref{Fig:part}.
Moreover, the horizon shrinks to zero size as the singularity is
approached since $\lim_{t_0 \rightarrow 0}\, r_h = 0$, and thus
asymptotic silence holds. However, if $\gamma \leq \frac{2}{3}$ no
particle horizon is formed, see Fig.~\ref{Fig:part}, and hence
asymptotic silence is broken.

In this example the Hubble-normalized spatial frame vectors are given by
\be
	E_\alpha{}^i = \frac{3\gamma}{2}\, t^{(3\gamma-2)/3\gamma}\, 
	{\rm diag}\,(1,1,1)\, ,
\ee
so that
\be
\lb{E_FL}
\lim_{t\rightarrow 0^+}E_\alpha{}^i =
\left\{\begin{array}{ll}
{\rm diag}\, (0,0,0) &,\quad \sfrac{2}{3} < \gamma \leq 2 \\
{\rm diag}\, (1,1,1) &,\quad \gamma = \sfrac{2}{3} \\
{\rm diag}\, (+\infty,+\infty,+\infty) &,\quad
0 < \gamma \leq \sfrac{2}{3}
\end{array}\right. \, ,
\ee
which shows that in this example the condition (\ref{E0}) is a reliable 
indicator of the existence of a particle horizon.
This result is to be compared with the casual properties depicted
in Fig.~\ref{Fig:part}.

\subsection{Asymptotic scalar curvature singularity
structure}\lb{sec:scss}

The second aspect of curvature singularities that is relevant to
our discussion is the asymptotic behaviour of the Ricci and Weyl
curvature scalars.

Firstly, it is of interest to consider the Hubble-normalized 
curvature scalars, the density parameter
\be
\lb{def_Om}
	\Om = \frac{\rho}{3H^2}\, ,
\ee
describing Hubble-normalized Ricci curvature (equivalently
$R_{ab}R^{ab}/H^4$), and
\be 
\lb{ccccs_def}
	\cc:=\frac{C_{abcd}C^{abcd}}{H^4}\, ,\quad
	\ccs:=\frac{C^*_{abcd}C^{abcd}}{H^4}\, , 
\ee
i.e, the Hubble-normalized Weyl scalar and pseudoscalar invariants. If 
$\lim_{\ell\rightarrow 0}\Om \neq 0$, then the Ricci scalar curvature, or 
equivalently, the matter content, is dynamically significant at the
singularity and if $\lim_{\ell\rightarrow 0}\cc\neq0$ or
$\lim_{\ell\rightarrow 0}\ccs\neq0$, then the Weyl scalar
curvature is dynamically significant at the singularity.

Secondly, it is of interest to consider the ratios
\be 
	\ccr:=\frac{C_{abcd}C^{abcd}}{\rho^2}\, ,\quad
	\ccrs:=\frac{C^*_{abcd}C^{abcd}}{\rho^2}\,  
\lb{def_ccr}
\ee
(or equivalently $C_{abcd}C^{abcd}/R_{ab}R^{ab}$ and
$C^*_{abcd}C^{abcd}/R_{ab}R^{ab}$), indicating the relative rates
at which the Ricci and Weyl curvature scalars diverge. 
Note that
\be
	\ccr  = \frac{\cc}{9\Omega^2}\, ,\quad
	\ccrs = \frac{\ccs}{9\Omega^2}\, ,
\lb{new_ccr}
\ee
as follows from (\ref{def_Om})--(\ref{def_ccr}). 
We here introduce the following classification.
 A scalar curvature
singularity is said to be 
\begin{itemize}
\item Ricci-dominated if $\lim_{\ell
\rightarrow 0}\ccr = 0$ and $\lim_{\ell \rightarrow 0}\ccrs=0$
\item Weyl-Ricci balanced if
$\ccr\,,\ccrs$ are bounded, and at least one of them
 has a non-zero limit, or oscillates indefinitely as $\ell \rightarrow 0$
\item Weyl-dominated if one of $\ccr\, ,\ccrs$ is
unbounded, possibly oscillating 
\end{itemize}
(see~\ct{barher02} for a similar
classification used for discussing future asymptotes, although
$C^*_{abcd}C^{abcd}$ was not considered in that paper).

As an example, consider an isotropic singularity (see e.g.,
\cite{goowai85}, \cite{limetal04}). In this case the
Hubble-normalized Ricci curvature is dynamically significant while
the Hubble-normalized Weyl curvature is not,
\be 
	\lim_{\ell \rightarrow 0} \Omega = 1\, ,\quad \lim_{\ell
	\rightarrow 0} \cc=0\, ,\quad \lim_{\ell \rightarrow 0} \ccs=0\, .
\ee
This leads to $\lim_{\ell \rightarrow 0} \ccr=0$ and $\lim_{\ell
\rightarrow 0} \ccrs=0$, and one thus has asymptotic Ricci
dominance. On the other hand, at a generalized Mixmaster
singularity (see \ct{uggetal03}), $\lim_{\ell \rightarrow 0} \Omega =0$ 
while
the Weyl curvature is dynamically significant, which yields
asymptotic Weyl dominance. Indeed, $\lim_{\ell \rightarrow 0} \cc$
and $\lim_{\ell \rightarrow 0} \ccs$ do not even exist, which
implies that $\lim_{\ell \rightarrow 0} \ccr$ and $\lim_{\ell
\rightarrow 0} \ccrs$ do not exist either. The non-existence of
these limits is due to the oscillating nature of the generalized
Mixmaster dynamics. However, $\cc$ and $\ccs$ are bounded (since
they are bounded on the Kasner subset and the vacuum Bianchi
type II subset which describe the attractor for Mixmaster
dynamics, see \ct{uggetal03}) and hence $\cc$ and $\ccs$ exhibit bounded
oscillations, while $\ccr$ and $\ccrs$ oscillate in an unbounded
manner.

\subsection{Asymptotic spatial structure}\lb{sec:sps}

In~\cite{waiell97} (p. 30) a classification of matter
singularities is given that reflects the change of shape of a
spherical element of a fluid as a singularity is approached. This
classification is based on the limiting behaviour of the scale
factors $\ell_\alpha$ in the eigendirections of the expansion
tensor, and is applicable to models in which the limits of the
scale factors exist or are infinite. In practice it is more
convenient to base this classification on the Hubble-normalized
expansion tensor (see e.g., \cite{waiell97} p. 19)
\be \Theta_{\alpha\beta} = \frac{\theta_{\alpha\beta}}{H}\, , \ee
which is related to the Hubble-normalized shear tensor%
\footnote{$\Sigma_{\alpha\beta}$ is defined by (\ref{Hnorm}) in the
Appendix.}
$\Sigma_{\alpha\beta}$ according to
\be \Theta_{\alpha\beta} = \delta_{\alpha\beta} +
\Sigma_{\alpha\beta}\, . \ee
It follows that $\Theta^\alpha{}_\alpha=3$. We are considering
models for which the limit $\Theta_{\alpha\beta}$ exists as the
singularity is approached, say
\be \lim_{\ell\rightarrow 0}\,\Theta_{\alpha\beta} =
\Theta^s_{\alpha\beta}\, , \ee
where $\Theta^s_{\alpha\beta}$ is a $3\times3$ symmetric matrix,
which in general depends on spatial position. We diagonalize
$\Theta^s_{\alpha\beta}$, and let $\Theta^s_{\alpha}$ denote its
diagonal entries. If the limit of the deceleration parameter
exists,
\be \lim_{\ell\rightarrow 0}\, q = q^s\, , \ee
then one can define the {\em shape parameters\/} $p_\alpha$
according to
\be 
\lb{shape}
	p_\alpha = \frac{\Theta^s_{\alpha}}{1 + q^s}\, ,\quad
{\textstyle{\sum\limits_{\alpha=1}^3}}\, p_\alpha =
\frac{3}{1+q^s}\, ,\ee
which for the well-known Kasner solution become the
so-called Kasner parameters. The exponents $p_\alpha$ are related
to diagonalized limit values of $\Sigma_{\alpha\beta}$ by
\be 
	\Sigma^s_{\alpha} = \frac{3p_\alpha}{\sum p_\alpha} - 1\, ,
\lb{Sig_p}
\ee

The classification (first introduced by Thorne~\cite{tho67} for
Bianchi type I cosmologies) is based on the signs of the
$p_{\alpha}$:

\begin{tabular}{lrl}
	& (i) & $(+,+,+)$, point, \\
	& (ii) & $(+,+,0)$, barrel,\\
	&(iii) & $(+,+,-)$, cigar, \\
	& (iv) & $(+,0,0)$, pancake,\\
\end{tabular}

\noindent
and cycle on 1,2,3, cf. p. 121~\ct{waiell97}. Heuristically, if
$\Theta^s_1>0$ ($=0,<0$, respectively), for example, the length
scale in the 1-direction will tend to zero at the singularity
(respectively a $const,+\infty$), thereby justifying this
terminology.

\subsection{Asymptotically self-similar singularities}
\lb{sec:sss}

The simplest cosmological singularities are \emph{asymptotically 
self-similar}, in the sense that the limits of all Hubble-normalized 
scalars exist as $\ell \rightarrow 0$, and equal the constant values of an 
exact self-similar SH solution, which we shall refer to as the 
\emph{asymptotic solution at the singularity}.
The examples in Section~\ref{sec:SHnon}, except for the 
Szekeres and
Wainwright-Marshman solutions in Sections~\ref{subsec:sze}
and \ref{subsec:waima}, have an 
asymptotically self-similar singularity. The corresponding asymptotic 
solutions are as follows:
\begin{itemize}
\item[i)]	the Taub form of flat spacetime \cite[p. 193]{waiell97},

\item[ii)]	the diagonal vacuum plane wave solution 
		\cite[p. 191]{waiell97},

\item[iii)]	the Collins VI$_h$ perfect fluid solution 
			\cite[p. 190]{waiell97}.
\end{itemize}
One can think of the Taub form of flat spacetime as the special Kasner 
vacuum solution for which the curvature tensor is zero. For brevity we 
shall refer to this form of flat spacetime as the \emph{Taub solution}.
Likewise we shall refer to ii) as the \emph{plane wave solution}.

For any self-similar SH solution, the Hubble-normalized expansion tensor 
$\Theta_{\alpha\beta}$ has constant components, and the deceleration 
parameter $q$ is constant. These quantities determine the shape 
parameters $p_\alpha$ according to (\ref{shape}). For the three 
above-mentioned solutions we have:
\begin{itemize}
\item[i)] The Taub solution:
	\be
	\lb{Tpp}
		(p_\alpha) = (1,0,0)\, .
	\ee

\item[ii)] The plane wave solution: 
	\be
		(p_\alpha) = (1,r+\sqrt{r(1-r)},r-\sqrt{r(1-r)}),
	\ee
	where $r$ satisfies $0 < r < 1$. The $p_\alpha$ satisfy 
	the conditions
	\be
	\lb{pwpp}
		p_1 + p_2 + p_3 = p_1^2 + p_2^2 + p_3^2 > 1\, .
	\ee

\item[iii)] The Collins VI$_h$ solution: 
	\be
	\lb{CVIhpp1}
		(p_\alpha) = (1,\sfrac{2-\gamma+rs}{2\gamma},
			\sfrac{2-\gamma-rs}{2\gamma}),
	\ee
	where $s^2=(2-\gamma)(3\gamma-2)$, with $\sfrac23 < \gamma < 2$,
	and $r$ satisfies $0 < r < 1$. The $p_\alpha$ satisfy
        the conditions
        \be
	\lb{CVIhpp}
		\sfrac{2}{\gamma} 
		= p_1 + p_2 + p_3 > p_1^2 + p_2^2 + p_3^2\, .
	\ee
\end{itemize}
The line element for the Taub solution is given by
\be
	ds^2 = -dt^2 + t^2 dx^2 + dy^2 + dz^2 \, .
\ee
Full details about the solutions ii) and iii) are given in Appendix B. 
For a given cosmological solution with asymptotically self-similar 
singularity, one can identify the asymptotic solution by calculating the 
$p_\alpha$.

\section{Explicit examples}
\lb{sec:SHnon}

We divide our examples into two main categories: spatially
homogeneous examples and inhomogeneous examples. The quantity
$\CCS$, and thus $\ccs$ and $\ccrs$, is identically zero for all 
examples except for the Wainwright-Marshman solutions in
Section~\ref{subsec:waima}. Hence these quantities will not be 
given except in this last case.

\subsection{Spatially homogeneous examples}
\lb{subsec:SH}

\subsubsection{Bianchi type I perfect fluid solution} 
\lb{subsubsec:LRS}

We consider the LRS Bianchi type I perfect fluid solution with a linear
equation of state, given by (\cite{waiell97}, p. 199 with
$p_\alpha=(1,0,0)$):
\be
	ds^2 = -A^{2(\gamma-1)}\,dt^2 + t^2\, A^{-\frac{2}{3}}\,dx^2
	+ A^{\frac{4}{3}}(dy^2+dz^2)\, ,
\lb{BI_ds}
\ee
where
\be
	A^{2-\gamma}= \alpha + m^2\,t^{2-\gamma }\, .
\ee
The matter quantities are 
\be
	{\bf u}=A^{-(\gamma-1)}\,\partial/\partial t\, ,\quad 
	\rho = \frac{4m^2}{3t^\gamma\,A^\gamma}\, ,\quad 
	p=(\gamma-1)\rho\, ,
\lb{BI_matter}
\ee
where $\alpha$ and $m$ are two positive constants, and $\gamma$ satisfies 
$0 < \gamma < 2$.

The Hubble scalar is given by
\be
	H = \sfrac13 A^{1-\gamma} t^{-1}
        	[ 1 + A^{-(2-\gamma)} m^2 t^{2-\gamma} ]
	\sim t^{-1} \rightarrow \infty\, ,
\lb{BI_H}
\ee
and the Weyl scalar by
\be
	\CC = \frac{64 \alpha^2 m^4}{27 A^4 t^{2\gamma}}\, .
\lb{BI_Weyl}
\ee
There is a curvature singularity at $t=0$, and the rates of growth of the 
curvature scalars are
\be
        \rho \sim t^{-\gamma} \rightarrow \infty\, , \quad
        \CC \sim t^{-2\gamma} \rightarrow +\infty\, ,
\lb{lrsc}
\ee
as follows from (\ref{BI_matter}) and (\ref{BI_Weyl}).
The shape parameters are $p_\alpha = (1,0,0)$, which by (\ref{Tpp}) shows 
that the asymptotic solution at the singularity is the Taub solution.
In addition it follows from Section~\ref{sec:scss} that the singularity
is of the pancake type.
Equations (\ref{BI_H}) and (\ref{lrsc}) imply that
the density parameter and the Hubble-normalized Weyl scalar satisfy
\be
	\Om \sim t^{2-\gamma} \rightarrow 0\, ,\quad 
	\cc \sim t^{2(2-\gamma)} \rightarrow 0\, ,
\ee
which is consistent with being asymptotic to the Taub
solution. 
In addition, it follows from (\ref{BI_matter}) and (\ref{BI_Weyl}) that 
the ratio $\ccr$, as defined by (\ref{def_ccr}), satisfies
\be
        \ccr \rightarrow \sfrac{4}{3}\, .
\lb{lrsccr}
\ee
Equations (\ref{lrsc})--(\ref{lrsccr}) show that the singularity is
Weyl-Ricci  
balanced, and is such that neither the Weyl nor the Ricci curvature is
dynamically significant.

\begin{figure}[h]
\begin{center}
\includegraphics{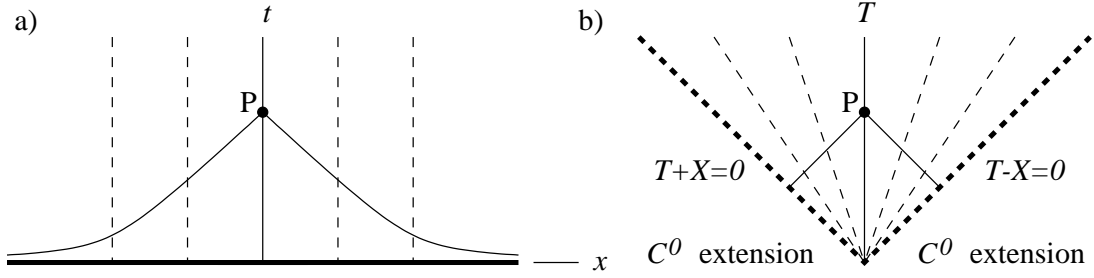}
\caption{Figures a) and b) show the causal structure associated
with the LRS Bianchi I solution that exhibits a weak null singularity, and
illustrates asymptotic silence-breaking. It is an example in which no
particle horizon forms in one direction, i.e., it is an example of
a partially silent singularity. Fig. \ref{Fig:LRS}a) utilizes the present
coordinates while Fig. \ref{Fig:LRS}b) uses coordinates adapted to the 
null structure of the singularity. In both figures the thin dashed lines
represent the fluid worldlines.} \label{Fig:LRS}
\end{center}
\end{figure}

It follows from (\ref{BI_ds}) and (\ref{BI_H}) that
the Hubble-normalized frame components satisfy
\be
	E_\alpha{}^i \sim {\rm diag}\,(\beta,t,t) \rightarrow
		(\beta,0,0)\, ,
\lb{lrse}
\ee
where $\beta = 3 \alpha^{\frac{3\gamma-2}{3(2-\gamma)}}$,
showing asymptotic silence-breaking in the $x$-direction, and this is 
further confirmed by the form of past null cone of a typical point, as 
shown in Fig.~\ref{Fig:LRS} a).

The dust case is discussed extensively
by Hawking and Ellis
 on p. 145-147 in \cite{hawell73}. 
By means of a coordinate transformation that takes the
Taub line element, $-dt^2 + t^ 2\,dx^2$, to explicit flat form 
$-dT^2+dX^2$, they show that one can make a $C^0$ extension of the metric
across a null surface where the above ``interior'' solution is joined
with flat spacetime 
(see our Fig.~\ref{Fig:LRS}b and Fig. 22 
on p.146 in~\cite{hawell73}). For this reason the singularity is referred 
to
as a ``weak null singularity''.

\subsubsection{Bianchi type VI$_h$ two-fluid solution}
\lb{subsubsec:wai}

Wainwright~\ct{wai83} has given an exact SH
perfect fluid solution of Bianchi type VI$_h$
 with a barotropic equation of state. For
our purposes it is useful to consider a special case of this
solution determined by setting $u=\frac{1}{10},v=w=\frac{3}{10}$
in~\cite{wai83}, in which case the group parameter is $h=-1/9$.
In this case the matter content can be
interpreted as two non-interacting dust and radiation fluids. The
line-element is given by:
\be 
	ds^2 = A^{-2}\,(-dt^2 + t^2\,dx^2) + (t\,e^x)^{4/5}\,dy^2 +
		A^{2}\,(t\,e^x)^{-2/5}\,dz^2\, ,
\lb{VI_ds}
\ee
where
\be
	A= \alpha_s + \alpha_m\,t^{2/5}\, .
\ee
The constants $\alpha_s,\alpha_m$ are required to satisfy $\alpha_s > 0, 
\alpha_m > 0$.
The matter quantities are given by
\be
	{\bf u}=A\,\partial/\partial t\, ,\quad
	\rho = \rho_R + \rho_M\, ,\quad 
        p = \sfrac13 \rho_R\, ,
\lb{VI_matter}
\ee
where
\be
	\rho_R = \sfrac{12}{25}\alpha_s\alpha_m\,t^{-8/5}\, ,\quad
	\rho_M=\sfrac{8}{25}\alpha_m^2\,t^{-6/5}\, .
\lb{VI_matter2}
\ee
Note that asymptotically the equation of state is linear with a radiation 
equation of state, i.e. $\lim_{t\rightarrow 0} p/\rho = 1/3$. 

The Hubble scalar is given by
\be
	H = \sfrac25 A t^{-1} \sim t^{-1} \rightarrow \infty\, ,
\lb{VI_H}
\ee
and the Weyl scalar by
\be
	\CC = - A^4 t^{-4} \left[ 2(\sfrac{24}{25})^2 T + O(T^2)
		\right]\, ,
\lb{VI_Weyl}
\ee
where $T = t \ptl_t \ln A \sim t^{2/5} \rightarrow 0$.
There is a curvature singularity at $t=0$, and the rates of growth of the
curvature scalars are
\be  
        \rho \sim t^{-8/5} \rightarrow \infty\, ,\quad
        \CC \sim - t^{-18/5} \rightarrow -\infty\, ,
\lb{6c}
\ee
as follows from (\ref{VI_matter}), (\ref{VI_matter2}) and (\ref{VI_Weyl}).
The shape parameters are given by
\be
        p_\alpha = (1,\sfrac{2}{5},-\sfrac{1}{5})\, ,
\ee
and satisfy (\ref{pwpp}), which shows that  the asymptotic solution at the 
singularity is the plane wave solution with $h=-\sfrac19$.
In addition, it follows from Section~\ref{sec:sps} that the singularity is 
of the cigar type.

Equations (\ref{VI_H}) and (\ref{6c}) imply
that
the density parameter and the Hubble-normalized Weyl scalar satisfy
\be
        \Om \sim t^{2/5} \rightarrow 0 \, ,\quad
        \cc \sim - t^{2/5} \rightarrow 0 \, ,
\ee
which is consistent with being asymptotic to the plane wave solution.
In addition, equation (\ref{6c}) shows that the ratio $\ccr$ satisfies
\be
        \ccr \sim -t^{-2/5} \rightarrow -\infty\, .
\lb{6ccr}
\ee
which implies that the singularity is Weyl-dominated, in the terminology 
of Section~\ref{sec:scss}.
   
It follows from (\ref{VI_ds}) and (\ref{VI_H}) that
the Hubble-normalized frame components satisfy
\be
	E_\alpha{}^i \sim {\rm diag}\,(const, t^{3/5}, t^{6/5})
		\rightarrow (const,0,0)\, ,
\lb{6e}
\ee
showing asymptotic silence-breaking in the $x$-direction.

It is instructive to introduce the same ``Taub'' coordinate 
transformation as was
used in \cite{hawell73} to show that the previous LRS Bianchi type I 
solution was $C^0$ extendible. Let
\be
\lb{transform}
\left\{
\begin{array}{l}
T+X = t\,e^x 
\\
T-X = t\,e^{-x}
\end{array}
\right.
\quad\Rightarrow 
\left\{
\begin{array}{l}
t = \sqrt{T^2-X^2} 
\\ 
e^x = \sqrt{\frac{T+X}{T-X}}
\end{array}
\right.\, ,
\ee
which leads to
\be
ds^2=A^{-2}\,(-dT^2 + dX^2) + (T+X)^{4/5}\,dy^2 + A^2\,(T+X)^{-2/5}\,dz^2\, , 
\ee
with $A=\alpha_s + \alpha_m\,(T^2-X^2)^{1/5}$. The metric is $C^0$ on 
$T-X=0$
where it can be joined with flat spacetime in the region $T-X<0$, but not 
on $T+X=0$.
Thus this plane wave singularity is also (partially) a weak null singularity.

\begin{figure}[h]
\begin{center}
\includegraphics{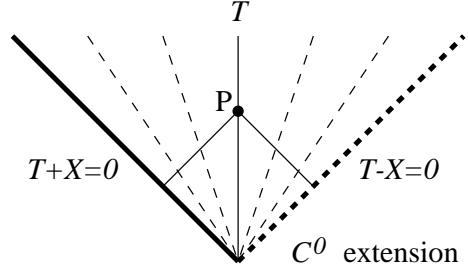}
\caption{The causal structure associated with the
Bianchi type VI$_h$ two-fluid solution in $T,X$ coordinates.}
\label{Fig:VIh}
\end{center}
\end{figure}

\subsection{Inhomogeneous examples}
\lb{subsec:inhom}

\subsubsection{Senovilla-Vera (1997) dust solution}
\lb{subsubsec:senver97}

Senovilla and Vera have presented a number of explicit dust solutions 
in~\ct{senver97}. One which is useful for our purposes
 is case (i) on p. 3483 in~\ct{senver97}. 
This solution can be written in fluid comoving coordinates so that 
the line-element takes the form:
\be
	ds^2 = -dt^2 + t^2\,dx^2 + Y^2\,dy^2 
		+ (t\,e^{-x})^{1-\rr}\,dz^2\, ,
\lb{SV_ds}
\ee
where
\bea
Y &=& c_-\,(t\,e^{-x})^{b_-} + m^2\,(t\,e^{x})^{b} + 
c_+\,(t\,e^{-x})^{b_+}\, ,\\
b&=&\sfrac{1}{2}(1+\rr)\, ,\quad b_\pm =
\sfrac{1}{2}(1\pm\sqrt{2-\rr^2})\, ,\quad -1<\rr<1\, . \eea
The parameters $b,b_\pm$ satisfy $b_- < 0 < b < 1 < b_+$. The
matter quantities are given by
\be
	{\bf u} =\partial/\partial t\, ,\quad
	\rho = \frac{(1-\rr^2)\,m^2\,(t\,e^x)^b}{t^2\,Y}\, ,\quad p=0\, .
\lb{SV_matter}
\ee
The Hubble scalar is given by
\be
	H = \frac{1}{6t} \left[ 3-\rr  + 2 Y^{-1}
	[ c_- b_-\,(t\,e^{-x})^{b_-} + m^2 b\,(t\,e^{x})^{b} +
	c_+ b_+\,(t\,e^{-x})^{b_+} ] \right]\, ,
\lb{SV_H}
\ee
and there is a simple expression for the Weyl scalar:
\be
	\CC = 
	\rho^2 \left[ \frac{4}{3} - \frac{2Y}{m^2\,(t\, e^x)}\right]\, .
\lb{SV_Weyl}
\ee

The solution yields three distinct types of singularity depending
on the sign of $c_-$. For our purposes, however, it suffices to consider
the cases $c_->0$ and $c_-=0$.

\

\paragraph{Case i)} $c_- > 0$

\

There is a curvature singularity at $t=0$, and the rates of growth of the 
curvature scalars are
\be
	\rho \sim t^{-2+b-b_-} \rightarrow \infty\, ,\quad
	\CC \sim - t^{-4+b-b_-} \rightarrow - \infty\, ,
\lb{case1_rhoCC}
\ee
as follows from (\ref{SV_matter}) and (\ref{SV_Weyl}). The shape 
parameters are given 
by
\be
	p_\alpha = (1,b_-,1-b)\, .
\ee
It follows that the $p_\alpha$ satisfy (\ref{pwpp}), which shows that the 
asymptotic solution at the singularity is the diagonal plane wave 
solution, with 
$ h = -[1-\frac12 (\rr+\sqrt{2-\rr^2})]/[1+\frac12 (\rr+\sqrt{2-\rr^2})]$.
Since $b_- < 0 < b < 1$, the singularity is of the cigar type.

It follows from (\ref{SV_H}) and (\ref{case1_rhoCC}) that
the density parameter and the Hubble-normalized Weyl scalar both tend to 
zero:
\be
	\Om \sim t^{b-b_-} \rightarrow 0 \, ,\quad 
	\cc \sim - t^{b-b_-} \rightarrow 0 \, ,
\ee
as required for a solution that is asymptotic to the 
diagonal plane wave solution.
In addition, equation (\ref{case1_rhoCC}) shows that the ratio $\ccr$ 
satisfies
\be
	\ccr \sim -t^{-(b-b_-)}\rightarrow -\infty\, ,
\ee
which implies that the singularity is Weyl-dominated.

This inhomogeneous solution is asymptotically homogeneous in the sense 
that the spatial Hubble gradient tends to zero:
\be
	r_1 \sim t^{b-b_-} \rightarrow 0\, .
\ee

\begin{figure}[h]
\begin{center}
\includegraphics{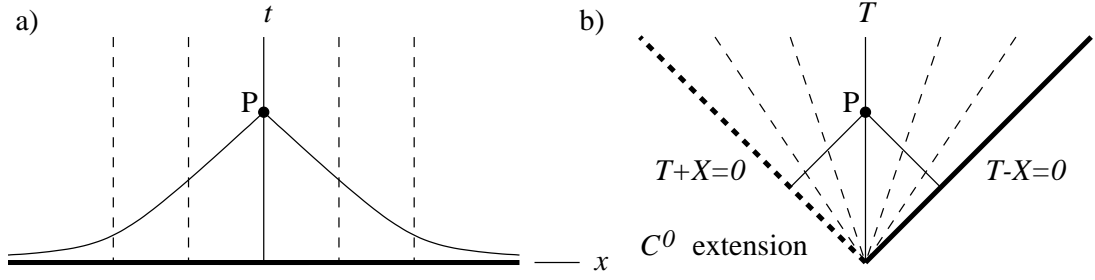}
\caption{a) shows the causal structure associated with the
Senovilla-Vera 1997 solution in both cases $c_- > 0$ and $c_- = 0$,
 in the original coordinates,
while b) shows it for the $c_->0$ case in $T,X$ coordinates.} 
\label{Fig:SV97}
\end{center}
\end{figure}

It follows from (\ref{SV_ds}) and (\ref{SV_H}) that
the Hubble-normalized frame components have the following limit:
\be
        E_\alpha{}^i \sim {\rm diag}\,\left( \frac{3}{2-b+b_-}\, ,\,
        t^{b_+}\, ,\, t^{b} \right) \rightarrow (const,0,0)\, ,
\ee
which suggests asymptotic silence-breaking in the $x$-direction, confirmed 
in Fig.~\ref{Fig:SV97}.

\

\paragraph{Case ii)} $c_- = 0$

\

There is a singularity at $t=0$, with $Y=0$ also, and the rates of growth 
of the curvature scalars are
\be
        \rho \sim t^{-2} \rightarrow \infty\, ,\quad
        \CC \sim -t^{-4} \rightarrow -\infty \, ,
\lb{case2_rhoCC}
\ee
as follows from (\ref{SV_matter}) and (\ref{SV_Weyl}).
The shape parameters are given by
\be
        p_\alpha = (1,\sfrac{1}{2}(1+\rr),\sfrac{1}{2}(1-\rr))\, .
\ee
It follows that the $p_\alpha$ are given by (\ref{CVIhpp1}) with 
$\gamma=1$ and $r=\rr$, which shows that the asymptotic solution at the 
singularity is the Collins VI$_h$ solution.
Since $-1 < k < 1$, the singularity is of the anisotropic point type.

It follows from (\ref{SV_H}) and (\ref{case2_rhoCC}) that
the density parameter and the Hubble-normalized Weyl scalar
have the following limits:
\be
        \Om \rightarrow \sfrac{3}{4}(1-\rr^2)\, ,\quad
        \cc \rightarrow - \sfrac{27}8 (1-\rr^2)^2\, ,
\lb{case2_Omcc}
\ee
as required for a solution that is asymptotic to the
Collins VI$_h$ solution for dust (set $\gamma=1$, $r=\rr$ in 
(\ref{CVI_Om}) and (\ref{CVI_Weyl})).
In addition, equations (\ref{case2_Omcc}) and (\ref{new_ccr}) show that
\be
        \ccr \rightarrow -\sfrac{2}{3}\, ,
\ee
which implies that the singularity is Weyl-Ricci balanced.

This solution is also asymptotically homogeneous, since
\be
	r_1 \sim t^{b_+-b} \rightarrow 0\, .
\ee

It follows from (\ref{SV_ds}) and (\ref{SV_H}) that
the Hubble-normalized frame components have the following limit:
\be
        E_\alpha{}^i \sim {\rm diag}\,\left(
        \sfrac{3}{2}\, ,\, t^{(1-\rr)/2}\, ,\, t^{(1+\rr)/2} \right)
        \rightarrow (const,0,0)\, ,
\ee
which shows asymptotic silence-breaking in the $x$-direction.

As in the SH cases it is instructive to 
make a ``Taub'' transformation of the coordinates (\ref{transform}). 
It turns out that
the $c_->0$ case is $C^0$ extendible through $T+X=0$, but not through $T-X=0$
(the $c_-=0$ case is not extendible) and thus the singularity in this 
case is (partially) a weak null singularity. See Fig.~\ref{Fig:SV97} 
for the causal structure of the singularity
for the various cases. Asymptotic silence is broken in
the $x$-direction and the singularity is partially silent for both cases.

\subsubsection{Szekeres solution} \lb{subsec:sze}

The Szekeres solutions~\ct{sze79}, \ct{goowai82} are a family of 
cosmological
dust solutions. These solutions can be interpreted as exact
perturbations of the Friedmann-Lema\^{\i}tre
 models, with both a growing 
and decaying
mode. For our purposes it suffices to consider a special Szekeres
model whose line-element is 
\be 
	ds^2 = -dt^2 + t^{4/3}\,( X^2 dx^2 + dy^2 + dz^2)\, ,
\lb{sze_ds}
\ee
where
\be
	X=a+k\,x\,t^{-1}\, ,
\ee
and $a,k$ are positive constants.
The matter quantities are given by
\be
	{\bf u} = \partial/\partial t\, ,\quad
	\rho = \sfrac{4}{3}\,a\,t^{-2}\,X^{-1}\, ,\quad
	p = 0\, .
\lb{sze_matter}
\ee

The Hubble scalar is given by
\be  
	H = \frac{2}{3t} \left[ 1 - \frac{k x}{2 t X} \right] \, ,
\lb{sze_H}
\ee
and the Weyl scalar by
\be
	\CC = \sfrac43 \rho^2 \left( 1-\frac{X}{a} \right)^2\, .
\lb{sze_Weyl}
\ee

If $x<0$ the singularity is determined by $X=0$, and the
rates of growth of the curvature scalars are
\be
        \rho \sim X^{-1} \rightarrow \infty\, ,\quad
        \CC \sim X^{-2} \rightarrow \infty\, ,
\lb{szec}
\ee
as follows from (\ref{sze_matter}) and (\ref{sze_Weyl}).
The shape parameters are $p_\alpha = (1,0,0)$, which 
shows that the singularity is of the pancake type.

It follows from (\ref{sze_H}) and (\ref{szec}) that
the density parameter and the Hubble-normalized Weyl scalar satisfy
\be
        \Om \sim X \rightarrow 0\, ,\quad
        \cc \sim X^2 \rightarrow 0\, ,
\ee
and in addition, equations (\ref{sze_matter}) and (\ref{sze_Weyl}) show 
that
\be
        \ccr \rightarrow \sfrac{4}{3}\, ,
\lb{szeccr}
\ee
which implies that the singularity is Weyl-Ricci balanced.

The preceding results suggest that the solution is asymptotic to the Taub 
solution at the singularity.
However, the behaviour of the spatial Hubble gradient $r_\alpha$ shows 
that the solution is not asymptotically spatially homogeneous.
It follows from (\ref{sze_H}) and (\ref{qr_alpha}) that there is one non-zero 
component given by
\be
	r_1 = \frac{3kat^{4/3}}{X(kx-2tX)^2}\, ,
\ee
which implies that
\be
	r_1 \sim X^{-1} \rightarrow \infty\, .
\ee
It thus appears that the asymptotic state is not adequately described by 
the Taub solution.
We shall refer to this singularity as being asymptotically ``Taub-like".

It follows from (\ref{sze_ds}) and (\ref{sze_H}) that
the Hubble-normalized frame components satisfy
\be
        E_\alpha{}^i \rightarrow (f(x),0,0)\, ,
\ee
where $f(x)$ is a positive function,
suggesting asymptotic silence-breaking in the $x$-direction.

\begin{figure}[h]
\begin{center}
\includegraphics{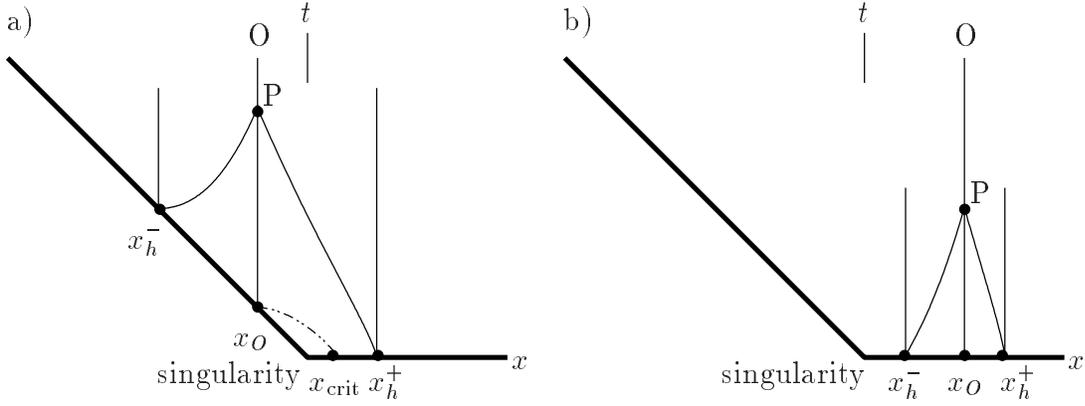}
\caption{This figure shows the particle horizon at an event P for
an observer O, generated numerically. The particle horizon is determined 
by the past null geodesics through P, which intersect the singularity at 
$x=x^+_h$ and $x=x^-_h$. In case a) the particle horizon does not shrink 
to
zero since as P approaches the singularity, $x^+_h$ approaches
$x_{\rm crit}$, the value determined by the dotted null geodesic.
In case b) the particle horizon does shrink to zero since
$x^{\pm}_h \rightarrow x_O$ as P approaches the singularity.} 
\label{Fig:sze}
\end{center}
\end{figure}

The causal structure in this example is quite different than in 
the previous examples.
Even though $E_1{}^1 \not \rightarrow 0$, a particle horizon does form 
for an observer with $x<0$, and inititally shrinks in size as $X 
\rightarrow 0^+$.
However, the horizon does not shrink to zero size, as is confirmed by the 
numerical simulations that are shown in Fig.~\ref{Fig:sze}.
 Thus asymptotic 
silence is broken for the $x<0$ part of the singularity. 

This example also illustrates that the nature of the singularity can 
depend on the fluid worldline.
If $x=0$ a two-parameter set of fluid lines approach an asymptotically
silent isotropic singularity (also known as an LK asymptote) 
\cite{goowai85}, \cite{limetal04}
characterized by $p_\alpha=\frac{2}{3}(1,1,1)$ and $\Om\rightarrow 1$. 
If $x>0$ the fluid lines approach
an asymptotically silent singularity of the cigar type described
by the LRS Kasner asymptote $p_\alpha = \frac{1}{3}(-1,2,2)$ (see Appendix 
B).

As a result, the limits of $\Om$ and $\cc$ at the singularity are 
position-dependent (i.e. depend on $x$). 
In particular
$\Omega\rightarrow 0$ when $x\neq 0$ but 
$\Omega\rightarrow 1$ at $x=0$, 
i.e. as the singularity is approached $\Om$ develops
a so-called spike
(see also~\ct[pp 91--96]{lim04}). 
In addition,
$\cc\rightarrow 0$ when $x\leq 0$ while $\cc\rightarrow const$ when $x>0$, 
i.e., $\cc$ asymptotically approaches a step function.

\subsubsection{Wainwright-Marshman solutions} \lb{subsec:waima}

The stiff perfect fluid solution found by Wainwright and
Marshman~\ct{waimar79} can be written as
\bea
 ds^2 &=& t^{2m}\,e^n\,(-dt^2 + dx^2) + t^{1/2}\,(dy +
w\,dz)^2 + t^{3/2}\,dz^2\, ,
\\
\rho &=& p = (m+\sfrac{3}{16})\,t^{-2(m+1)}\,e^{-n}\, ,\quad 
{\bf u}=t^{-m}\,e^{-n/2}\,\partial/\partial t\, . 
\lb{WM_matter}
\eea
Here $w=w(t-x)$ is an arbitrary function, $n=n(t-x)$ is
determined according to
\be n^\prime = (w^\prime)^2\, , \ee
and $m$ is a constant that satisfies $m\geq -\frac{3}{16}$.

The Hubble scalar is given by
\be
	H = t^{-(m+1)}\, e^{-n/2}\, B\, ,
\ee
where
\be
	B = \sfrac{1}{3}\,[(m+1)+\sfrac{1}{2}\,t\,(w')^2]\, .
\ee
The non-zero Hubble-normalized components of the spatial frame vectors are
\be
	E_1{}^1 = \frac{t}{B}\, ,\quad
	E_A{}^I = \frac{t^{m+\frac{1}{4}}\,e^{n/2}}{B}\, \left(
	\begin{array}{cc}
	    t^{1/2} & 0
	    \\
	    -w & 1
	    \end{array} \right)\, .
\ee
The Hubble-normalized kinematic quantities are
\be
	\Sig_{11} = 2 - \frac{1}{B}\, ,\quad
	\Sig_{22} = \frac{1}{4B} -1\, ,\quad
	\Sig_{33} = \frac{3}{4B} - 1\, ,\quad
	\Sig_{23} = \frac{t^{1/2}\,w'}{2B}\, ,
	\Udot_1 = -\frac{t\,(w')^2}{2B}\, .
\lb{wm_vars}
\ee
Finally 
the deceleration parameter 
and
the non-zero component of 
the spatial Hubble gradient 
are
\be
	q = -1+ \frac{m+1}{B} - \frac{t\,(w')^2}{6B^2} - r_1\, ,\quad
	r_1 = -\frac{t\,(w')^2}{2B} + \frac{t^2\,w'\, w''}{3B^2}\, .
\lb{wm_qr}
\ee

The Weyl tensor can best be described by giving the complex Newman-Penrose 
scalars \ct{newpen62}
\bea
\psi_0 &=& -\sfrac{1}{4}\, m\,t^{-(m+1)}\,e^{-n}\, ,\\
\psi_2 &=& \sfrac{1}{8}\,[\sfrac{4}{3}\,(m-\sfrac{3}{8}) +
i\,w'\,t^{1/2}]\,t^{-(m+1)}\,e^{-n}\, ,\\
\psi_4 &=& [-\sfrac{1}{4}\,m + \sfrac{3}{4}\,(w')^2\,t + i(w'' -
(w')^3 - m\,w'\,t^{-1})t^{3/2}]\,t^{-(m+1)}\,e^{-n}\,  \eea
(see \ct{wai86}, p. 3037 for these formulae).
The Weyl curvature scalars are then given by
\be C_{abcd}C^{abcd} - i\,C^*_{abcd}C^{abcd} = 16\,(3\psi_2^2 -
4\psi_1\,\psi_3 + \psi_0\,\psi_4)\, . \ee

We choose the arbitrary function $w(t-x)$ so that $n(t-x)$ satisfies
\be
\lb{cond1} 
	\lim_{t-x\rightarrow 0^+}\, n(t-x) = -\,\infty\, . 
\ee
The spacetime region is then defined by the inequalities
\be
\lb{cond2} 
	t>0\, ,\qquad t-x > 0\, .
\ee
Fluid worldlines with $x\leq0$ encounter a curvature
singularity at $t=0$ and those with $x>0$, at $t=x$, since $\rho
\rightarrow \infty$ in both cases, as follows from (\ref{WM_matter}).
In other words, the given initial singularity
has two branches that depends on the sign of $x$, namely $t=0,x \leq 0$ 
and $t=x, x>0$, respectively. 

\begin{figure}[h]
\begin{center}
\includegraphics{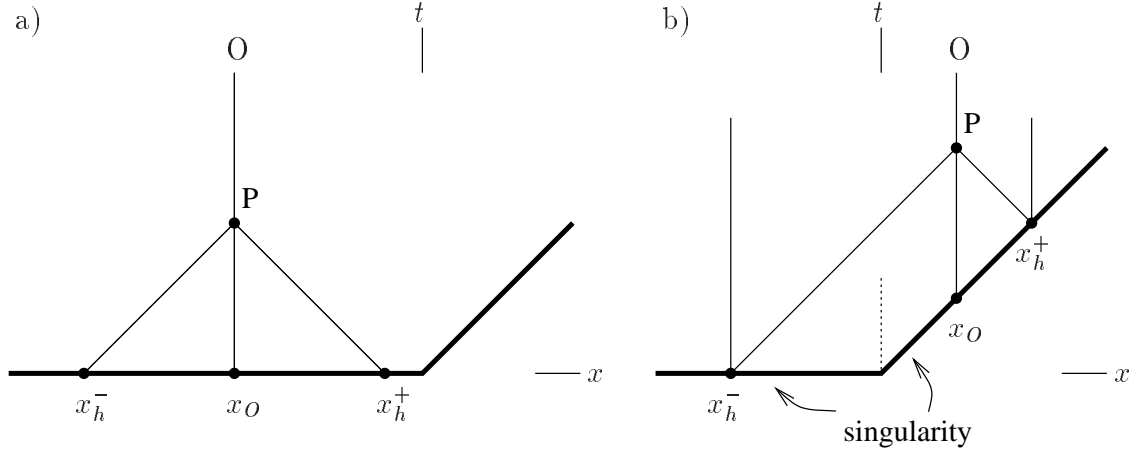}
\caption{The figure shows the causal structure associated with the
Wainwright-Marshman solution. The left part of the singularity is
asymptotically silent and is described asymptotically by the 
Jacobs stiff fluid solution. The right part of the singularity breaks
asymptotic silence.}
\label{Fig:WM}
\end{center}
\end{figure}

At the $t=0, x\leq 0$ part of the singularity the solution satisfies
\be \lim_{t\rightarrow 0^+}\, r_1 = 0 = \lim_{t\rightarrow 0^+}
\Udot_1\, ,\quad \lim_{t\rightarrow 0^+}\, (\Sig_{\alpha\beta}) =
\frac{1}{m+1}\,{\rm diag}\, (2m-1\, ,\, -\sfrac{1}{4}\,(4m+1)\,
	,\, \sfrac{1}{4}\,(5 - 4m))\, , 
\ee
which implies that the shape parameters $p_\alpha$ are given by
\be 
	p_\alpha = \frac{1}{m+1}\, (m, \sfrac{1}{4},\sfrac{3}{4})\, ,
\ee
and satisfy (\ref{KJ}).
The solution is thus asymptotic to a Jacobs stiff fluid solution.
The singularity type is: cigar $(-\frac{3}{16} <m<0)$, barrel
$(m=0)$, or anisotropic point $(m>0)$.
We also have
\be
	\lim_{t \rightarrow 0^+} E_\alpha{}^i =0,
\ee
and so we expect that the singularity is asymptotically silent, which is 
confirmed by Fig.~\ref{Fig:WM}a).

The nature of the $t-x=0, x>0$ part of the singularity depends on the
behaviour of the arbitrary function $w(t-x)$ or $n(t-x)$. 
However, since the null geodesics in the $tx$-space are given by $t \pm 
x=const$,
the causal structure in the $tx$-space is independent of the nature of the 
behaviour of $w(t-x)$ or $n(t-x)$.
Observe that the particle horizons of observers approaching the part of 
the singularity given by $t-x=0, x>0$ do not shrink to zero size as the 
singularity is approached. 
Hence asymptotic silence is broken.
Even though the causal structure is unaffected, the curvature properties 
depend significantly on the choice of $w(t-x)$ and $n(t-x)$.

There are two possibilities that are compatible with
equation~(\ref{cond1}):
\begin{itemize}
\item[(i)] $\lim_{t\rightarrow x^+}\, n'(t-x) = +\infty$
\item[(ii)]$n'(t-x)$ oscillates and is unbounded as $t\rightarrow
x^+$.
\end{itemize}

\paragraph{Case (i):} $\lim_{t\rightarrow x^+}\, n'(t-x) = +\infty$

\

It follows that
\be
	\lim_{t\rightarrow x^+} B = + \infty\, \quad 
	(\text{recall $n' = (w')^2$})\, , 
\ee
which in turn implies
\be
	\lim_{t\rightarrow x^+} \Sig_{\alpha\beta} =
	\text{diag}(2,-1,-1)\, ,\quad
	\lim_{t\rightarrow x^+} q = 2\, .
\ee
It follows that the shape parameters are $p_\alpha = (1,0,0)$, suggesting 
that the solution is asymptotic to the Taub solution.
But equations (\ref{wm_vars}) and (\ref{wm_qr}) imply that%
\footnote{Here we make the additional assumption that
$
	\lim_{t\rightarrow x^+} n''(t-x)/[n'(t-x)]^2=0
$,
which will be satisfied if, for example, $n(t-x)=-C(t-x)^{-b}$, where $b$ and $C$
are positive constants.}
\be 
	\lim_{t\rightarrow x^+} r_1 = \lim_{t\rightarrow x^+} \Udot_1
	= -3\, ,
\ee
so that the asymptotic state is inhomogeneous.
It thus appears that the asymptotic state is not adequately described by 
the Taub solution.
We shall refer to this singularity as being asymptotically ``Taub-like" (See
Section~\ref{subsec:sze}).

In addition,
we have the following asymptotic expressions:
\bea 
	H &\sim& e^{-n/2} (w')^2 \rightarrow \infty\, ,
\\
	\rho &\sim& e^{-n} \rightarrow \infty\, ,\quad
	\CC \sim e^{-2n} (w')^2 \rightarrow \infty\, ,
\\
	\Om &\sim& (w')^{-4} \rightarrow 0\, ,\quad
	\cc \sim (w')^{-6} \rightarrow 0\, ,
\\
	\ccr &\sim& (w')^2 \rightarrow \infty\, .
\eea
This part of the singularity is thus a Weyl-dominated scalar
curvature singularity.
Although it is ``Taub-like" as regards the shape parameters, the curvature ratio
$\ccr$ does not have the same limit at the singularity as in the solution that is
asymptotic to the Taub solution (see equation (\ref{lrsccr})). It is also
worth
noting that $\Om \rightarrow 0$, which shows that this solution is not in the 
generic class of stiff fluid solutions which satisfy $\Om \not\rightarrow 0$, 
and whose behaviour near the singularity was analyzed by Andersson and 
Rendall \ct{andren01}.

Even though $E_\alpha{}^i \rightarrow 0$ as the singularity is
approached, asymptotic silence is still broken, because the particle 
horizon to the left of an observer with $x>0$ does not shrink to zero at 
the singularity, i.e. $x_h^-$ does not tend to $x_O$ in 
Fig.~\ref{Fig:WM}b).
Thus this example
shows that $E_\alpha{}^i \rightarrow 0$ {\em does not necessarily lead
to horizons that shrink to zero\/}. This example thus forces us to
make a distinction between $E_\alpha{}^i \rightarrow 0$ and
particle horizons that shrink to zero. Since asymptotic silence
should reflect an invariant physical causal property we therefore
choose to define asymptotic silence as horizons shrinking to zero,
cf.~\cite{andetal05}. However, it should be pointed out that although
the solution approaches the so-called silent
boundary (see \ct{uggetal03} and \cite{andetal05}), it approaches 
a part not described by SH equations (see \ct{uggetal03} and 
\cite{andetal04} for a
discussion about the equations on the silent boundary) since
$\Udot_\alpha , r_\alpha \neq 0$. It is likely that this affects the 
relationship between $E_\alpha{}^i$ and the causal structure.

\

\paragraph{Case (ii):} $n'(t-x)$ oscillates and is unbounded as 
			$t\rightarrow x^+$

\

Let us consider the following possible choice~\ct{wai86}:
\bea w'(t-x) &=& \frac{\alpha}{t-x}\,\sin\,\left( \frac{b}{t-x}
\right)\, ,
\quad t-x>0\, \quad \Rightarrow \\
n(t-x) &=& -\,\frac{\alpha^2}{2(t-x)} + \frac{\alpha^2}{4b}\,
\sin\,\left(\frac{2b}{t-x} \right)\, \eea
(so that $\lim_{t\rightarrow x^+} n = -\infty$, as required in
equation~(\ref{cond1})). This yields
\begin{itemize}
\item $\lim_{t\rightarrow x^+} E_A{}^I =0$ ($A,I=2,3$), while
 $\lim_{t\rightarrow x^+} E_1{}^1$ does not
exist, since $E_1{}^1$ oscillate while remaining bounded.
It follows that the singularity is non-silent. 
\item $\Om$ oscillates and is bounded.
\item $\lim_{t\rightarrow x^+} \Sig_{\alpha\beta}$ does 
not exist because of (bounded) oscillations. 
\item $\cc$ oscillates and is bounded but $\ccs$ oscillates and is 
{\em unbounded\/} (due to $w''$). 
\end{itemize}

It is perhaps natural to refer to this kind of singularity as
being strongly Weyl-dominated, 
i.e. \emph{the Hubble-normalized Weyl curvature is unbounded}.%
\footnote{This behaviour at the singularity is
reminiscent of the future evolution of Bianchi type
VII$_0$ models (see~\ct{waietal99}).}
This solution is thus a counterexample to the hope that the 
Hubble-normalized Weyl 
curvature is bounded at any cosmological singularity. 
It is nevertheless possible that
the Hubble-normalized Weyl scalars are bounded at
asymptotically silent singularities.

\section{Concluding remarks}
\lb{sec:conc}

In this paper we have presented a collection of cosmological
solutions that illustrate the phenomenon of asymptotic silence-breaking.
The examples, apart from the Szekeres and Wainwright-Marshman
solutions, 
are asymptotically self-similar at the singularity, in the sense that the 
limits of all Hubble-normalized variables exist as $\ell \rightarrow 0$,
and equal the constant values of an exact self-similar SH solution.
The properties of the singularities of these solutions are
summarized in Table~\ref{tab:selfsim}.

\begin{table}
\begin{tabular}{ccccc}
Solution & asymptotic solution & asymptotic spatial structure
& $\lim\limits_{\ell \rightarrow 0} \ccr$ 
& $\lim\limits_{\ell \rightarrow 0} q$
\\
\hline\\
LRS Bianchi type I & Taub & pancake & $\sfrac{4}{3}$ & 2
\\ \\
Bianchi type VI$_h$ & plane wave ($h = -\sfrac19$)
& cigar & $-\infty$ & $\sfrac32$
\\
two-fluid
\\ \\
SV (1997) dust $c_->0$ & plane wave ($-1 < h < 0$)
& cigar & $-\infty$ & $\sfrac{2}{1-3h}$
\\ \\
SV (1997) dust $c_-=0$ & Collins VI$_h$ ($-1 < h < 0$)
& point & $-\sfrac{2}{3}$
& $\sfrac12$
\\ \\
Szekeres & Taub-like & pancake & $\sfrac{4}{3}$ & 2
\\ \\
\hline
\end{tabular}
\caption{Summary of properties of the diagonal
solutions that exhibit asymptotic silence-breaking at the singularity.}
\label{tab:selfsim}
\end{table}

We comment on a number of common features of these examples:
\begin{itemize}
\item[i)]
	All of the solutions have a Ricci and a Weyl curvature 
	scalar singularity in the sense that
	\be
		\lim_{\ell \rightarrow 0} \rho = \infty\, ,\quad
		\lim_{\ell \rightarrow 0} \CC = \pm \infty\, .
	\lb{rhoCC_infty}
	\ee
\item[ii)]
	Asymptotic silence-breaking occurs in one direction only.
\end{itemize}
On the other hand, the differences are found in:
\begin{itemize}
\item[i)]
	The Hubble-normalized Weyl curvature components and scalars, and 
	the density parameter:
	\begin{xalignat*}{4}
	&\text{Taub} 
	&& 
	\ce_{\alpha\beta}\ \&\ \ch_{\alpha\beta} \rightarrow 0
	&&
	\cc \rightarrow 0
	&&
	\Om \rightarrow 0
\\
	&\text{Plane wave}   
        &&
        \ce_{\alpha\beta}\ \&\ \ch_{\alpha\beta} \not\rightarrow 0
        &&
        \cc \rightarrow 0
        &&
        \Om \rightarrow 0
\\
	&\text{Collins VI$_h$}   
        &&
        \ce_{\alpha\beta}\ \&\ \ch_{\alpha\beta} \not\rightarrow 0
        &&
        \cc \not\rightarrow 0
        &&
        \Om \not\rightarrow 0
	\end{xalignat*}
	The variables $\ce_{\alpha\beta}$ and $\ch_{\alpha\beta}$ are
	Hubble-normalized frame components of the electric and magnetic parts of
	the Weyl tensor as defined in Appendix A, equation (\ref{cech_def}).
\item[ii)]
	Asymptotic spatial structure (see Table~\ref{tab:selfsim})
\item[iii)]
	Weyl-Ricci balance, as described by the ratio $\ccr$ (see 
	Table~\ref{tab:selfsim})	
\end{itemize}

It is of interest to ask what statements can be made in general about 
solutions that are asymptotic to the Taub solution, the plane wave 
solution and the Collins VI$_h$ solution at the singularity.
On the basis of the properties of the asymptotic solutions in Appendix B, 
we can draw the following conclusions:
\begin{itemize}
\item	In any perfect fluid solution asymptotic to the Taub solution at 
	the singularity,
	asymptotic silence-breaking occurs with rank $(E_\alpha{}^i) = 1$, 
	for any $\gamma$, $0 < \gamma < 2$. The spatial singularity type 
	is pancake.
	We are unable to predict whether 
		$\lim_{\ell \rightarrow 0} \ccr = \frac43$ 
	in general, since the ratio $\ccr$ is undefined for the Taub 
	solution.
	A detailed analysis of the asymptotic behaviour would be needed to 
	answer this question.

\item	In any SH perfect fluid solution asymptotic to the plane wave 
	solution at the singularity,
	if $1 \leq \gamma < 2$ then
	asymptotic silence-breaking occurs with rank $(E_\alpha{}^i) = 1$,
	and the singularity is of the cigar type.
	The group parameter $h$ will satisfy $-1 < h < 0$.
	We are unable to predict whether
		$\lim_{\ell \rightarrow 0} \ccr = -\infty$
	in general (if $\gamma \geq 1$) since the ratio $\ccr$ is 
	undefined for the plane wave solution.
	A detailed analysis of the asymptotic behaviour would be needed to
        answer this question.

\item	In any perfect fluid solution asymptotic to the Collins VI$_h$ 
	solution at the singularity,
	if \mbox{$1 \leq \gamma < 2$} then
	asymptotic silence-breaking occurs with rank $(E_\alpha{}^i) = 1$.
	The group parameter $h$ will satisfy $-1 \leq h < 0$.
	The spatial singularity type is point, barrel or cigar, and is 
	necessarily a point if $\gamma=1$.
	We can also predict that if $\gamma=1$, the limit of the curvature 
	ratio $\ccr$ will be $-\frac23$.
	If $1 < \gamma < 2$, the limit will also be non-zero, but its 
	value will depend on $\gamma$ and $h$ through equations
	(\ref{CVI_Om}) and (\ref{CVI_Weyl}).

\end{itemize}

The remaining examples, the Szekeres solutions and the Wainwright-Marshman
solutions (case ii)), are not asymptotically self-similar at the
asymptotically silence-breaking part of the singularity. 
In the Szekeres solution, asymptotic self-similarity fails because
$\lim_{\ell \rightarrow 0^+} r_1 = +\infty$, while the behaviour of the
Hubble-normalized curvature scalars ($\Omega$, $\cc$ and $\ccr$) is the same as
the solutions that are asymptotic to the Taub solution. For this reason we refer
to the Szekeres solution as being ``Taub-like" at the singularity.
In the Wainwright-Marshman solutions of class ii), the failure of
asymptotic self-similarity is reflected in the curvature,
in the sense that the limits of the Hubble-normalized energy
density and Weyl tensor as $\ell \rightarrow 0$ do not exist.
Indeed, $\ccs$ is even {\em unbounded\/}. They also
constitute an interesting example of asymptotic silence-breaking,
since $E_1{}^1$ oscillates so that the limit of this quantity does not
exist. 

The Wainwright-Marshman solutions that belong to class (i)
exhibit asymptotic silence-breaking even though $E_\alpha{}^i\rightarrow 
0$.
In this solution, however, $r_\alpha,\Udot_\alpha$ have non-zero 
limits. 
This raises the question: are the conditions 
$\lim_{\ell \rightarrow 0}\,(E_\alpha{}^i,r_\alpha,\Udot_\alpha) = 
(0,0,0)$,
given by equations (\ref{E0}) and (\ref{Ur0}),
sufficient for asymptotic silence,%
\footnote{They are not necessary conditions since recurring spike
formation (see \cite{andetal05}) provides a counterexample; in this
case,
asymptotic
silience holds, with $\lim_{\ell \rightarrow 0} E_\alpha{}^i =0$, but
$\lim_{\ell \rightarrow 0} r_\alpha$ and
$\lim_{\ell \rightarrow 0} \Udot_\alpha$ do not exist.}
i.e., the existence of particle horizons
that shrink to zero as the singularity is approached?

\

\noindent
{\bf Acknowledgement}

\ 

\noindent
WCL and CU gratefully acknowledge the Isaac Newton Institute For
Mathematical Sciences at the University of Cambridge
 where part of this work was done.
CU is supported by the Swedish Research Council.
JW is supported by the Natural Sciences and Engineering Research Council
of Canada.
\

\appendix

\section{Orthonormal frame formalism and Hubble-normalized variables}
\lb{sec:def}

In this paper we use the orthonormal frame formalism in conjunction with 
Hubble-normalized variables to calculate the properties of the explicit 
solutions. 
We introduce an orthonormal frame $\{ \vece_0, \vece_\alpha \}$, where 
$\vece_0 = \mathbf{u}$, the 4-velocity of the perfect fluid, which is 
assumed to be irrotational. Local coordinates $t,x^i$ are introduced such 
that
\be
	\vece_0 = N^{-1} \ptl_t\, ,\quad
	\vece_\alpha = e_\alpha{}^i \ptl_i\, ,
\ee
where $\ptl_t$, $\ptl_i$ denote partial derivatives with respect to $t$ 
and $x^i$ ($\alpha = 1,2,3$, $i=1,2,3$).

The commutators are decomposed according to:
\bea \lb{comts0} [\,\vece_{0}, \vece_{\alpha}\,]\,(f) & = &
\dot{u}_{\alpha}\,\vece_{0}(f) - [\,H\,\d_{\alpha}{}^{\beta} +
\sig_{\alpha}{}^{\beta} +
\eps_{\alpha}{}^{\beta}{}_{\gam}\,\Om^{\gam}\,]\,
\vece_{\beta}(f) \\
\lb{comtsa} [\,\vece_{\alpha}, \vece_{\beta}\,]\,(f) & = &
(2a_{[\alpha}\,\d_{\beta]}{}^{\gam} +
\eps_{\alpha\beta\delta}\,n^{\delta\gam})\,\vece_{\gam}(f) \ .
\eea
Here $H$ is the Hubble scalar, $\udot^{\alpha}$ the acceleration,
$\sig_{\alpha\beta}$ the tracefree shear, $\Om^{\alpha}$ describes
the angular velocity of the spatial frame $\{\,\vece_{\alpha}\,\}$
relative to a frame that is Fermi-propagated along the integral curves 
of $\vece_{0}$, while $a^{\alpha}$ and
$n_{\alpha\beta}=n_{(\alpha\beta)}$ determine a spatial
connection. Scale-invariant (dimensionless) Hubble-normalized
variables are defined as follows (see \ct{uggetal03}; for a slightly
different set of definitions based on conformal considerations,
see \cite{rohugg05}):
\bea \lb{dlfram} & & \parb_{0} := \frac{\vece_{0}}{H} =
\cn^{-1}\,\ptl_{t} \ , \quad
\parb_{\alpha} := \frac{\vece_{\alpha}}{H} =
E_{\alpha}{}^{i}\,\ptl_{i}\ ,\quad \cn:= NH \ , \quad
E_{\alpha}{}^{i} := \frac{e_{\alpha}{}^{i}}{H} \ , \\
\lb{dlcon} & & \{\Udot^{\alpha}, \,\Sig_{\alpha\beta},
\,A^{\alpha}, \,N_{\alpha\beta}, \,R^{\alpha}\} : =
\{\udot^{\alpha}, \,\sig_{\alpha\beta}, \,a^{\alpha},
\,n_{\alpha\beta}, \,\Om^{\alpha}\}/H \, .
\lb{Hnorm}
\eea
It is convenient to express the Weyl tensor in terms of its electric and 
magnetic parts, relative to the timelike congruence $\vece_0$, denoted by
$E_{\alpha\beta}$ and $H_{\alpha\beta}$ (see \cite{waiell97}, p. 19).
Their 
Hubble-normalized components are defined by
\be
	(\ce_{\alpha\beta},\ch_{\alpha\beta})
	= (E_{\alpha\beta},H_{\alpha\beta})/(3H^2)\, .
\lb{cech_def}
\ee
The Weyl scalars introduced in Section~\ref{sec:scss} are expressed in 
terms of $E_{\alpha\beta}$ and $H_{\alpha\beta}$ as follows:
\be
	\CC = 8(E_{\alpha\beta} E^{\alpha\beta} 
		- H_{\alpha\beta} H^{\alpha\beta})\, ,\quad
	\CCS = 16 E_{\alpha\beta} H^{\alpha\beta}\, .
\ee
It follows from (\ref{ccccs_def}) and (\ref{cech_def}) that
the Hubble-normalized Weyl scalars are given by
\be
	\cc = 72(\ce_{\alpha\beta} \ce^{\alpha\beta}
		- \ch_{\alpha\beta} \ch^{\alpha\beta})\, ,\quad
	\ccs = 144 \ce_{\alpha\beta} \ch^{\alpha\beta} \, .
\lb{cc_EH}
\ee
We note that the components $E_{\alpha\beta}$, $H_{\alpha\beta}$ can be 
calculated using the orthonormal frame formulae in \ct{waiell97} (p 35). 
Similar formulae for the Hubble-normalized components $\ce_{\alpha\beta}$, 
$\ch_{\alpha\beta}$ are given in \ct{uggetal03} (p 21).

In terms of the orthonormal frame formalism,
the deceleration parameter
$q$ and the
frame components of
the  spatial Hubble gradient $r_a$, defined by (\ref{qr_a}), are given by
\be
\lb{qr_alpha} 
	\parb_{0}H = -\,(q+1)\,H \ , \quad
	\parb_{\alpha}H = -\,r_{\alpha}\,H \ , 
\ee
It follows from Raychaudhuri's equation (see \ct{uggetal03}) that
\be
	q = 2\Sig^{2} + \sfrac{1}{2}(3\gamma-2)\Om
	- \sfrac{1}{3}\,(\parb_{\alpha}-r_{\alpha}+\Udot_{\alpha}
	-2A_{\alpha})\,\Udot^{\alpha} \, ,
\ee
where $\Sig^{2}=\sfrac{1}{6}\,\Sig_{\alpha\beta}\Sig^{\alpha\beta}$.
The SH spacetimes are characterized by $\parb_\alpha (\ ) =0$ for all 
Hubble-normalized variables, and $r_\alpha = 0 = \Udot_\alpha$, with the
$\vece_\alpha$ being tangential to the SH symmetry surfaces.

\section{The asymptotic solutions}
\lb{sec:H4}

The asymptotic solutions that were introduced in Section~\ref{sec:sss} are 
non-tilted SH self-similar solutions with perfect fluid source. A complete 
list of these solutions (line-element and matter variables) is given in 
\cite{waiell97} (Section 9.1).
In this Appendix we give the Hubble-normalized variables for the specific 
solutions that are used in this paper,
which are the ones with diagonal line-element.
Since the line-elements are diagonal,  one can introduce an 
orthonormal frame in a natural way. For the line-element
\be
	ds^2 = - N^2 dt^2 + X^2 dx^2 + Y^2 dy^2 + Z^2 dz^2
\ee
we choose
\be
	\vece_0 = N^{-1} \ptl_t\, ,\quad
	\vece_1 = X^{-1} \ptl_x\, ,\quad
        \vece_2 = Y^{-1} \ptl_y\, ,\quad
        \vece_3 = Z^{-1} \ptl_z\, ,
\ee
so that the non-zero
components $e_\alpha{}^i$ of the spatial frame vectors are
\be
	e_1{}^1 = X^{-1}\, ,\quad
        e_2{}^2 = Y^{-1}\, ,\quad
        e_3{}^3 = Z^{-1}\, .
\ee
The self-similar solutions in question are either of Bianchi type I or 
type VI$_h$.

\subsection{Bianchi type I solutions}

The line-element is \cite[p. 187]{waiell97}
\be
	ds^2 = -dt^2 + t^{2p_1} dx^2 + t^{2p_2} dy^2 + t^{2p_3} dz^2\, .
\ee
The Hubble scalar and deceleration parameter are
\be
	H = b^{-1} t^{-1}\, ,\quad
	q = b -1\, ,\quad
	b = \sfrac3{p_1 + p_2 + p_3}\, ,
\ee
and the only non-zero Hubble-normalized commutation functions are
the diagonal shear components,
\be
	\Sigma_{\alpha\alpha} = b p_\alpha -1 \quad
\text{(no sum).}
\ee
The constants $p_\alpha$ are the shape parameters (see (\ref{Sig_p})).

\vspace*{3mm}

\noindent
(1) {\em Flat Friedmann-Lema\^{\i}tre solution}
\be 
\lb{flatFL} 
	p_1=p_2=p_3=\frac{2}{3\gamma}\, .
\ee
The matter density is given by
\be
	\rho = \frac{4}{3\gamma^2} t^{-2}\, ,\quad
	0 < \gamma \leq 2\, ,
\ee
which implies that the density parameter is $\Omega=1$.
The Weyl tensor $C_{abcd}$ is identically zero.

\vspace*{3mm}

\noindent (2) {\em Kasner vacuum solutions and Jacobs stiff perfect fluid
solutions}

\be
	p_1+p_2+p_3=1\, ,\quad 
	p_1^2+p_2^2+p_3^2 
		\begin{cases}
			= 1 &\text{for Kasner vacuum solutions,}
		\\
			< 1 &\text{for Jacobs stiff fluid solutions.}
		\end{cases}
\lb{KJ}
\ee
The non-zero Hubble-normalized Weyl tensor components are
\be
	\ce_{11} = 2p_2\,p_3 - p_1(p_2+p_3)\, .
\ee
where $\ce_{\alpha\beta} = {\rm diag}(\ce_{11},\ce_{22},\ce_{33})$
and where $\ce_{22}, \ce_{33},$ are obtained from $\ce_{11}$ by cyclically
permuting 1, 2, 3, and the density parameter is 
\be
	\Om = \sfrac32(1-p_1^2-p_2^2-p_3^2)\, .
\ee
The choice $(p_\alpha)=(1,0,0)$, and cycle, yields the Taub form of flat 
spacetime.

\subsection{Diagonal Bianchi type VI solutions}

The line-element is \cite[p. 190]{waiell97}
\be
	ds^2 = - dt^2 + t^2 dx^2 + t^{2 p_2} e^{2 c_2 x} dy^2
		+  t^{2 p_3} e^{2 c_3 x} dz^2 \, .
\ee
The Hubble scalar and deceleration parameter are
\be
	H = b^{-1} t^{-1}\, ,\quad
	q = b - 1\, ,\quad
	b = \sfrac3{1+p_2 + p_3}\, ,
\lb{DVI_H}
\ee
and the non-zero Hubble-normalized commutation functions are
\be
	\Sigma_{\alpha\beta} = \text{diag}(b-1,
	b p_2-1,b p_3-1)\, ,\quad
	A_1 = -\sfrac{b}{2}(c_2+c_3) \, ,\quad
	N_{23} = -\sfrac{b}{2}(c_2-c_3) \, .
\ee
The constants $p_1=1$, $p_2$, $p_3$ are thus the shape parameters (see
(\ref{Sig_p})).
The Hubble-normalized frame components are given by
\be
	E_1{}^1 = b \, ,\quad
	E_2{}^2 = b t^{1-p_2} e^{-c_2 x} \, ,\quad
	E_3{}^3 = b t^{1-p_3} e^{-c_3 x} \, .
\lb{DVI_Eai}
\ee

The non-zero Hubble-normalized Weyl tensor components are
\be
	\ce_{\alpha\beta}={\rm diag}(\alpha,-{\sfrac{1}{2}}\alpha+\beta,
		-{\sfrac{1}{2}}\alpha-\beta)\, ,\quad
	\ch_{23}=\sfrac1{12} b^2
		[(c_2+c_3)(p_2-p_3)-(c_2-c_3)(2-p_2-p_3)]\, ,
\lb{DVI_EH}
\ee
where
\be
	\alpha= \tfrac{1}{18}b^2 \left[ (c_2-c_3)^2 - (p_2-p_3)^2 \right]
\, ,\quad
	\beta = \tfrac{1}{12}b^2 \left[ (c_2^2-c_3^2) - (p_2-p_3)(2-p_2-p_3)
		\right]
\, .
\ee

\vspace*{3mm}

\noindent (1) {\em Collins Bianchi type VI$_h$ perfect fluid solution}

\be 
	p_{2,3}=\frac{2-\gamma\pm rs}{2\gamma}\, ,\quad 
	c_{2,3}=\frac{(2-\gamma)r \pm s}{2\gamma}\, ,\quad 
	s=\sqrt{(2-\gamma)(3\gamma-2)}\, .
\lb{CVI_pcs}
\ee
The parameter $r$ determines the group parameter $h$ according to
\be
	r^2=r_c^2\,(-h)\, ,\quad 
	r_c=\sqrt{(3\gamma-2)/(2-\gamma)}\, ,\quad
	0 \leq r < 1 \, .
\lb{CVI_r}
\ee
The matter quantities are given by
\be
	\rho = \frac{(2-\gamma)(1-r^2)}{\gamma^2} t^{-2}\, ,\quad
	\sfrac{2}{3} < \gamma < \sfrac{2(1-h)}{1-3h}\, ,
\ee
which implies, using (\ref{DVI_H}), that
\be 
	\Om= \tfrac{3}{4}(2-\gamma)(1-r^2)\, .
\lb{CVI_Om}
\ee
The Hubble-normalized Weyl scalar is given by
\be
	\cc = - \sfrac{27}{8} s^2(1-r^2)
	\left[ (4-3\gamma)(2-\gamma)(1-r^2) + 12 \gamma (\gamma-1) \right]
	\, ,
\lb{CVI_Weyl}
\ee
as follows from (\ref{DVI_EH}) and (\ref{cc_EH}).

The spatial singularity type depends on $\gamma$ and $h$. We restrict to 
non-negative pressure ($\gamma \geq 1$) which implies via (\ref{CVI_r})
that $r_c^2 \geq 1$ and hence that $-h \leq r^2$. On writing the shape 
parameters $p_2$ and $p_3$ in the form
\be
	p_{2,3} = \frac{2-\gamma}{2\gamma}(1 \pm r_c r)
\ee
using (\ref{CVI_pcs}) and (\ref{CVI_r}), we arrive at the classification 
shown in 
Table~\ref{tab:coll}.
It follows from (\ref{DVI_Eai}) that in all cases,
\be
	\lim_{t \rightarrow 0} E_\alpha{}^i =\text{diag}(const,0,0).
\ee

\begin{table}
\begin{tabular}{cccc}
parameter range & $p_\alpha$ 
& spatial singularity type & restriction on $\gamma$
\\
\hline
$ r r_c < 1$ & $(+,+,+)$ & anisotropic point & $1 \leq \gamma < 2$
\\ \\
$ r r_c = 1$ & $(+,+,0)$ & barrel & $1 < \gamma < 2$
\\ \\
$ r r_c > 1$ & $(+,+,-)$ & cigar & $1 < \gamma < 2$
\\
\hline
\end{tabular}
\caption{Classification of the singularity in the Collins Bianchi type
VI$_h$ perfect fluid solution with $\gamma \geq 1$.
The value $\gamma=1$ corresponds to $r_c=1$.}
\label{tab:coll}
\end{table}

\vspace*{3mm}

\noindent (2) {\em Diagonal vacuum plane wave solution}

\be 
	p_{2,3}=c_{2,3}=r \pm \sqrt{r(1-r)}\, ,
\lb{dpw_pc}
\ee
where $r$ is related to the group parameter $h$ through
\be
	h=-\frac{r}{1-r}\, .
\ee
The Hubble-normalized quantities are given by
\be 
	\Sig_{11} = \frac{2(1-r)}{1+2r}\, ,\quad
	\Sig_{22,33} = b\,p_{2,3} - 1\, ,\quad
	N_{23} = - b\,\sqrt{r(1-r)}\, ,\quad 
	A_1 = - b\,r\, .
\ee
The non-zero Weyl tensor components are
\be
	\ce_{22} = -\ce_{33} = \ch_{23} 
	= \sfrac13 b^2 (2r-1)\sqrt{r(1-r)}\, ,
\ee
showing that the Weyl scalars (\ref{ccccs_def}), as given by (\ref{cc_EH}), are
zero.

The spatial singularity type depends on $h$ (or equivalently $r$). The 
parameter $r$ is in turn restricted by the value of the equation of state 
parameter $\gamma$ of any non-tilted perfect fluid SH solution that is 
past asymptotic to the plane wave solution, according to
\be
	\gamma < \frac{2}{1+2r}
\ee
(see \cite[p. 663]{wai84}).
We restrict to $\gamma \geq 1$, which implies that $0 \leq r < \frac12$
(since $r \geq \frac12$ implies $\gamma < 1$).
It follows from (\ref{dpw_pc}) that $p_2 > 0$ and $p_3 < 0$, so that
the singularity is of the cigar type. It also follows from (\ref{DVI_Eai})
that
\be
        \lim_{t \rightarrow 0} E_\alpha{}^i =\text{diag}(const,0,0).
\ee



\end{document}